\documentclass[aps,reprint,prl,twocolumn,floats,floatfix,longbibliography,superscriptaddress]{revtex4-1}

\usepackage{amsmath}
\usepackage{amsfonts}
\usepackage{amssymb}
\usepackage{graphicx}
\usepackage{color}
\usepackage[colorlinks=true,citecolor=blue,linkcolor=blue,urlcolor=blue]{hyperref}
\usepackage{times}
\usepackage{amstext}
\usepackage{latexsym}
\usepackage[running,mathlines]{lineno}
\usepackage{verbatim}

\providecommand{\ket}[1]{\lvert #1 \rangle}

\begin{document}
\raggedbottom

\title{Continuous generation of quantum light from a single ground-state atom in an optical cavity}

\author{Celso Jorge Villas-Boas}
\affiliation{Departamento de F\'{i}sica, Universidade Federal de S\~{a}o Carlos, 13565-905 S\~{a}o Carlos, S\~{a}o Paulo, Brazil}

\author{Karl Nicolas Tolazzi}
\affiliation{Max-Planck-Institut f\"ur Quantenoptik, Hans-Kopfermann-Str.\ 1, D-85748 Garching, Germany}

\author{Bo Wang}
\email{bo.wang@mpq.mpg.de }

\affiliation{Max-Planck-Institut f\"ur Quantenoptik, Hans-Kopfermann-Str.\ 1, D-85748 Garching, Germany}

\author{Christopher Ianzano}
\affiliation{Max-Planck-Institut f\"ur Quantenoptik, Hans-Kopfermann-Str.\ 1, D-85748 Garching, Germany}

\author{Gerhard Rempe}
\affiliation{Max-Planck-Institut f\"ur Quantenoptik, Hans-Kopfermann-Str.\ 1, D-85748 Garching, Germany}

\begin{abstract}
We show an optical wave-mixing scheme that generates quantum light by means of a single three-level atom. The atom couples to an optical cavity and two laser fields that together drive a cycling current within the atom. Weak driving in combination with strong atom-cavity coupling induces transitions in a harmonic ladder of dark states, accompanied by single-photon emission via a quantum Zeno effect and suppression of atomic excitation via quantum interference. For strong driving, the system can generate coherent or Schr\"odinger cat-like fields with frequencies distinct from those of the applied lasers. 

\end{abstract}

\maketitle

The laser was, immediately after its invention, considered a solution seeking a problem \cite{Maiman1964}. Since then, however, the laser has opened up a plethora of possibilities that led to many scientific and technological advances. They all build on the fact that in a macroscopic optical resonator many excited atoms can emit many photons that form a bright but essentially classical field \cite{Schawlow1958}. The situation changes for a microscopic resonator where a single atom and a single photon can interact strongly to create a quantum-mechanical atom-photon molecule. This forms the basis for the research field of cavity quantum electrodynamics (QED) \cite{Haroche2013, Ritsch2013, Xiang2013} that has found novel applications in quantum-information processing \cite{Reiserer2015, Lodahl2015, Chang2018}. From a basic perspective, however, cavity QED remains a radically new and still unexplored platform for the control of light-matter interaction with possibilities that go far beyond those of a laser \cite{Dubin2010}.

Here we report on a new phenomenon that allows one to generate different quantum light fields with just one optically controlled atom in its ground state. To this end, we integrate cavity QED with the phenomenon of electromagnetically induced transparency (EIT) \cite{Alzetta1976, Harris1990, Fleischhauer2005, Mucke2010, Kampschulte2010, Albert2011, Souza2013} and predict continuous light emission without atomic excitation. Specifically, we employ an atom in a $\Lambda$-type three-level configuration (one excited and two ground states) where one branch is strongly coupled to an optical cavity and the other to a control laser. In the EIT regime, the atom remains in a state known as a dark state in which the atom does not absorb light due to destructive interference of excitation amplitudes. By introducing a second laser field that couples the two ground states, it is possible to drive a closed cycle within the energy-level scheme of the atom. As expected for several waves interacting with an optically nonlinear atom, this gives rise to a new radiation field via an optical wave-mixing process \cite{Dudin2012, Maser2016, Ripka2018}.

Not expected, however, is that a weak coupling laser does not destroy the fragile dark states of the cavity EIT system, even when all fields are on resonance with the respective atomic transitions. We then find that the two lasers, in combination with the cavity, drive optical transitions in a harmonic ladder of dark states that differ by one photon in the cavity. The dark nature of these states suppresses atomic excitation, but the entanglement with the cavity field introduces photons. For a weak coupling laser, the photons are produced one by one due to the presence of a continuous quantum Zeno effect \cite{Misra1977}. When increasing the intensity of the coupling laser, the system is perturbed and moves to a regime where the transitions to the excited state become strongly detuned. This detuning restricts the system dynamics to the ground states of the atom, and now leads to a coherent cavity field with uncorrelated photons. Remarkably, in all coupling cases the new field comes from a single atom in its ground state.

\begin{figure}[h]
	\includegraphics[width=8.3cm]{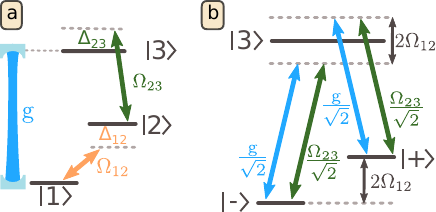}
	\caption{\label {fig_lvlscheme} Energy-level diagrams. (a) shows the level scheme of the system in the bare state basis where the driving strengths $\Omega_{12}$ and $\Omega_{23}$, the detunings $\Delta_{12}$ and $\Delta_{23}$, and the atom-cavity coupling strength $g$ are depicted. (b) shows the same system in the dressed state basis with new effective coupling strengths.}
\end{figure}
\begin{figure}[h]
	\includegraphics[width=8.3cm]{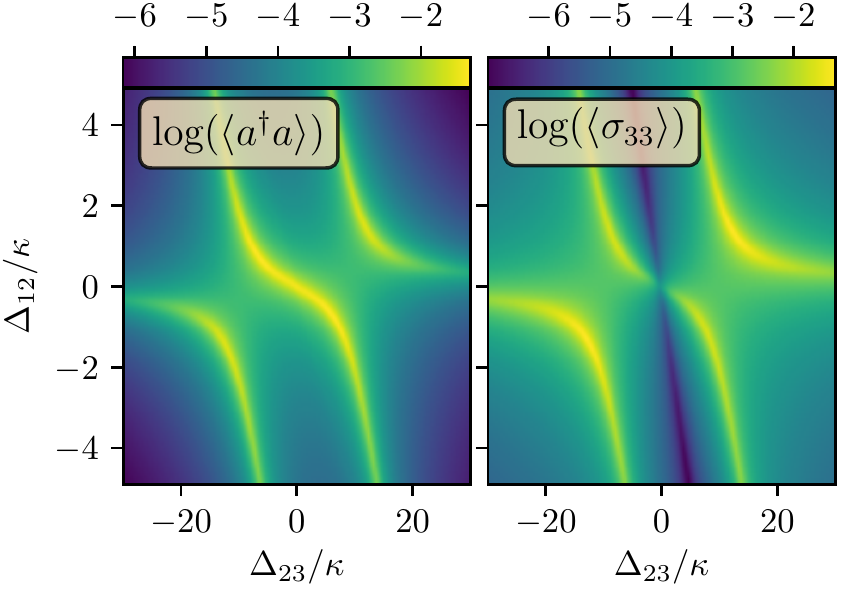}
	\caption{\label {fig_pops}Steady state photon number and atomic excited-state population. Shown are two-dimensional color plots of the respective observable as indicated in the upper left corner of the plot against $\Delta_{12}$ and $\Delta_{23}$. All parameters are given in terms of the cavity-field decay rate $\kappa/2\pi=1.$ The other parameters are: $g=10\kappa$, $\Omega_{23}=3\kappa, \Omega_{12}=0.1\kappa$, and $\Gamma_{31}=\Gamma_{32}=0.5\kappa$.}
\end{figure}

Our system consists of an effective three-level atom driven by two classical fields and strongly coupled to a quantum cavity mode, as schematically shown in Fig. \ref{fig_lvlscheme}(a). In this model, the ground states $\left|1\right\rangle$ and $\left|2\right\rangle$ are coupled to the excited state $\left|3\right\rangle$ via either the cavity mode, with coupling strength $g$ and frequency $\omega$, or the classical (control) field, with Rabi frequency and frequency $2\Omega_{23}$ and $\omega_{C23}$, respectively. We also consider another classical field coupling the $\left|1\right\rangle$ and $\left|2\right\rangle$ ground states, with Rabi frequency $2\Omega_{12}$ and frequency $\omega_{C12}$ (here dubbed ground-state coupling field). The Rabi frequencies, field frequencies, and atom-cavity coupling factors are all taken to be real and non-negative. The Hamiltonian of this system, in the interaction picture, is given by ($\hbar=1$) 
\begin{equation}
H=H_{0}+V,\label{hamiltonian}
\end{equation}
with $H_{0}=\left(ga\sigma_{31}+\Omega_{23}\sigma_{32}+\text{h.c.}\right)-\left(\Delta-\Delta_{12}-\Delta_{23} \right) a^{\dagger}a \\+\Delta_{23}\sigma_{33} -\Delta_{12}\sigma_{11} $ and $V=\Omega_{12}\sigma_{21}+\text{H.c.}$. Here $a$ ($a^{\dagger}$) is the annihilation (creation) operator of the cavity mode, $\sigma_{kl}=\left|k\right\rangle \langle l|$ ($k,l=1,2,3$) the atomic operators which describe the transition from the $\left|l\right\rangle $ to $\left|k\right\rangle$ states, and $\text{H.c.}$ stands for Hermitian conjugate. The detunings are given by $\Delta_{12}=\omega_{21}-\omega_{C12}$, $\Delta_{23}=\omega_{32}-\omega_{C23}$, and $\Delta=\omega_{31}-\omega$, with $\omega_{kl}$ the respective atomic transition frequency. The dynamics of our system are governed by the master equation $\dot{\rho}=-i[H,\rho]+\kappa(2a\rho a^{\dagger}-a^{\dagger}a\rho-\rho a^{\dagger}a)+\sum_{m=1,2}\Gamma_{3m}(2\sigma_{m3}\rho\sigma_{3m}-\sigma_{33}\rho-\rho\sigma_{33})$, where $\kappa$ is the cavity-field decay rate, and $\Gamma_{31}$ and $\Gamma_{32}$ the polarization decay rates of the excited level $|3\rangle$ to levels $|1\rangle$ and $|2\rangle$, respectively. The master equation is solved numerically using QuTip \cite{Johansson2013}. 

In Fig. \ref{fig_pops} we plot the average number of intracavity photons $\langle a^{\dagger}a\rangle$ and the population of the excited atomic state $\langle \sigma_{33} \rangle$ in steady state as a function of the detunings.  Three states and two avoided crossings can be observed in both the excited state and the cavity populations. Two of these states we designate bright states (see Supplemental Material \cite{supplement} for more information)  because they contain contributions from the atomic excited state $\ket{3}$, while the other is a dark state with only ground-state atomic contributions. A two-photon condition $\omega_{C12}+\omega_{C23}=\omega_{31}$ can be observed where the atomic excited-state population remains low.  This can be seen as a dark line in Fig. \ref{fig_pops}. For most values of $\omega_{C12}$ and $\omega_{C23}$ satisfying this two-photon condition, the cavity population is also negligible. However, for the situation in which all fields are resonant ($\Delta_{12}=\Delta_{23}=0$), the system produces photons in the cavity mode while the atomic population in the excited state $\ket{3}$ remains low. Thus, despite resonant driving, the atomic excitation is suppressed while photons are injected into the cavity. 


In order to understand the effect and explore its consequences, we proceed in two steps: we first present a simplified picture and then provide a full quantum-mechanical explanation for the two cases of weak and strong ground-state coupling. The simplified picture starts from the observation that for vanishing detunings the Hamiltonian, expressed in a new atomic basis $\{\left|+\right\rangle ,\left|-\right\rangle \}$ with $\left|\pm\right\rangle=\left(\left|1\right\rangle \pm\left|2\right\rangle \right)/\sqrt{2}$, can be understood as two independent $\Lambda$ schemes. This is displayed in Fig. \ref{fig_lvlscheme}(b) where two cavity EIT configurations \cite{Mucke2010} are shown (see Supplemental Material \cite{supplement}  for details) for which an interference process can be expected. This is in fact the case in the limit of very weak ground-state coupling [$\Omega_{12}\ll\kappa<(g,\Omega_{23})$], i.e. when the two states $\left| +\right\rangle$ and $\left| -\right\rangle$ are degenerate. In this case both the cavity and control field remain almost resonant with the atomic transitions $\left| +\right\rangle \leftrightarrow \left| 3 \right\rangle $ and $\left| -\right\rangle \leftrightarrow \left| 3 \right\rangle $, respectively. Interference of the absorption paths then avoids the excitation of the atom, keeping it always in the subspace of the dark states which involve only the atomic states $\left| +\right\rangle$ and $\left| -\right\rangle$.
In the strong ground-state coupling limit [$\Omega_{12}>(g,\Omega_{23})\gg\kappa$] the degeneracy of the two states $\left| +\right\rangle$ and $\left| -\right\rangle$ is lifted and the system is no longer resonantly driven. This regime can then be well described by two separate processes involving only a classically driven two-level atom coupled to a cavity mode. We now present the full quantum description for the two considered cases. 

\textit{Weak ground-state coupling regime.} For $(g,\Omega_{23})>\kappa\gg\Omega_{12}$, we treat the coupling field as a perturbation. In this case, the effective Hamiltonian, written in terms of the dark states of the unperturbed Hamiltonian $H_0$, reads (see Supplemental Material\cite{supplement} )
\begin{equation}
H_{\text{eff}}\simeq - \Omega_{12}\sum_{n=0}^{\infty}Q_n R_{n+1} \left|\Psi_{n+1}^{0}\right\rangle \left\langle\Psi_{n}^{0}\right|+\text{H.c.},
\label{Heff_RWA}
\end{equation}
with $Q_{n}=\Omega_{23}/\sqrt{g^2n+\Omega_{23}^2}$,  $R_{n}=g\sqrt{n}/\sqrt{g^2n+\Omega_{23}^2}$, and 
the dark states (for zero eigenvalues) \cite{Souza2013}
\begin{align}
\left\vert \Psi_{0}^{0}\right\rangle &=\left\vert 1,0\right\rangle \label{dark_state_0}\\
\left\vert \Psi_{n}^{0}\right\rangle  &=   Q_n\left\vert 1,n\right\rangle -R_n\left\vert 2,n-1\right\rangle.\label{dark_states}
\end{align}
Here, $n=1,2,...$ denotes the photon number in the cavity. The remarkable feature of this Hamiltonian is that it describes transitions only between dark states, and that bright states are not involved. Starting from the lowest dark state $\left\vert \Psi_{0}^{0}\right\rangle$ the system is coherently driven along the harmonic ladder of dark states with a rate that is proportional to the ground-state coupling constant $\Omega_{12}$, thereby changing the cavity excitation quanta by quanta. In this process the system remains always in the subspace of dark states. As these have no overlap with the atomic excited state, $\left|3\right\rangle$, its population remains negligible. This result is particularly important since it implies that any atomic decay does not influence the dynamics of the system, thereby eliminating problems such as the decay to other atomic states besides the relevant ones considered here. 

In the regime $g \gg\Omega_{23}$ the effective Hamiltonian can be further simplified, with the result 
\begin{equation}
H_{\text{eff}}\simeq-\Omega_{12}\left|\Psi_{1}^{0}\right\rangle \left\langle\Psi_{0}^{0}\right|-\sum_{n=1}^{\infty}\frac{\Omega_{12}\Omega_{23}}{g\sqrt{n}}\left|\Psi_{n+1}^{0}\right\rangle \left\langle\Psi_{n}^{0}\right|+\text{H.c.}
\label{Heff_weak_coupling}
\end{equation}
Obviously, the transition strength between dark states decreases nonlinearly with increasing $n$ so that transitions are mainly driven between the first two dark states, i.e., between the states which contain zero and one photon only. This has important consequences for the steady-state population distribution of the dark states. To understand which of these states contribute most to the dynamics of the system we also need to calculate their decay rates which can be derived via Fermi's golden rule (see Supplemental Material \cite{supplement} ):
\begin{equation}
\Gamma_{n}^0=\left|\langle\Psi_{n-1}^{0}\left|\sqrt{\kappa}a\right|\Psi_{n}^{0}\rangle\right|^{2} = n\kappa \left[ \frac{\Omega_{23}^2 + g^2(n-1)}{\Omega_{23}^2 + g^2n} \right] 
\end{equation}
Notice that the higher dark states decay only into lower ones since the cavity dissipation affects only the photon number and therefore cannot induce transitions from dark states to bright states \cite{Souza2013}. For $g\gg \Omega_{23}$ the decay rates of the dark states $|\Psi_{1}^{0}\rangle$ and $|\Psi_{2}^{0}\rangle$ reduce to $\Gamma_1^0 \simeq \kappa \left( \Omega_{23}/g \right)^2$ and $\Gamma_2^0 \simeq \kappa$, respectively. Thus, in this limit, the suppression of higher-photon number states due to the nonlinear driving [$\sim 1/\sqrt n$, see Eq. (\ref{Heff_weak_coupling})], together with the fact that dark states with more photons decay faster than the single-photon dark state, implies that these higher-lying states remain largely unpopulated.

\begin{figure}[h]
	\includegraphics[width=8.6cm]{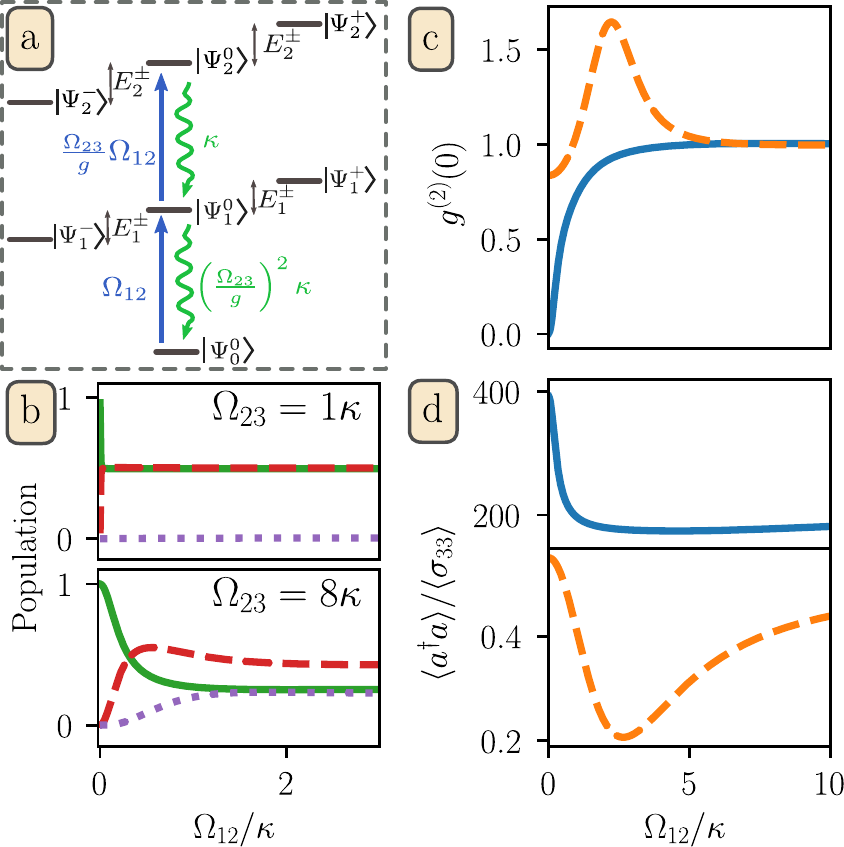}
	\caption{\label{fig_regimes} The sketch in a) shows the level scheme and the effective dark state driving strength and decay constant for weak ground-state coupling. Plot b) shows the population of the first few dark states $\left|\Psi_{0,1,2}^{0}\right\rangle$ (green solid, red dashed, purple dotted) against $\Omega_{12}$ for different $\Omega_{23}$ ($g=10\kappa$). The plots c) and d) show photon statistics and the relative photon number, respectively, against the coupling strength of the ground states $\ket{1}$ and $\ket{2}$ for two different values of the atom-cavity coupling strength $g$ (blue solid line $g=10\kappa$, orange dashed line $g=0.5\kappa$). The other parameters are $\Omega_{23}=3\kappa$ and $\Gamma_{31}=\Gamma_{32}=0.5\kappa$.  }
\end{figure}

The situation is depicted for the lowest three dark states in Fig. \ref{fig_regimes}(a), and simulation results are plotted in part (b) of the figure. It displays the steady-state populations $P_n^0$ of the dark states $\left|\Psi_{n}^{0}\right\rangle$ as a function of the normalized coupling strength $\Omega_{12}/\kappa$. For small $\Omega_{23}$, as shown in the upper plot, the population is equally distributed between $\left|\Psi_{0}^{0}\right\rangle$ and $\left|\Psi_{1}^{0}\right\rangle$, and the system can be well described as a two-level system with these two dark states. Only when $\Omega_{23}$ is comparable to the cavity coupling $g$, as depicted in the lower plot, higher dark states become populated and the two-level character is lifted, as further discussed below.

The finding that a harmonic ladder of states effectively reduces to the lowest two states can be interpreted differently: as the effective driving strength of the higher dark states ($n>1$) is much smaller than the cavity decay rate, $\kappa$, the atom-cavity system is constantly projected into the subspace spanned by $\ket{\Psi_0^0}$ and $\ket{\Psi_1^0}$ and therefore experiences a continuous quantum Zeno effect \cite{Misra1977}, that blocks access to higher rungs of the dark state ladder.

The two-level character of the system as predicted by the Zeno blockade has important consequences for the nature of the emitted light. This is demonstrated in Fig. \ref{fig_regimes}(c) which features the equal-time photon correlation function $g^{(2)}(0)=\langle a^{\dagger 2}a^2 \rangle/\langle a^{\dagger}a \rangle^2$. It goes to zero in the limit of weak driving ($\Omega_{12} \rightarrow 0$) and strong atom-cavity coupling $g$ (blue solid line). It follows that photons are emitted one after the other. The effect is nontrivial and also predicts, e.g., Rabi oscillations between the two lowest dark states, even for vanishing cavity decay (see Fig. S3 from the Supplemental Material\cite{supplement} ). On the other hand, for weak atom-cavity coupling, $g^{(2)}(0)$ is close to $1$ even in the limit of $\Omega_{12} \rightarrow 0$, as shown by the orange dashed line in Fig. \ref{fig_regimes}(c). In this case, the coupling field can induce transitions to higher dark states, and also to bright states, since the energy difference between the dark and the bright states ($E^{\pm}_{n}=\pm\sqrt{ng^2+\Omega^2_{23}} $) is small, see Fig. \ref{fig_regimes}(a).

The figure of merit of the excitationless emission is emphasized in Fig. \ref{fig_regimes}(d) which demonstrates that photons are created in the cavity while avoiding the excited atomic state. The effect is characterized by the ratio $\langle a^{\dagger}a \rangle /\langle \sigma_{33} \rangle$ which can be much larger than one. It is important to stress that this occurs only in the strong atom-cavity coupling regime: for small values of $g$ the atom is strongly excited, thus resulting in a ratio $\langle a^{\dagger}a \rangle /\langle \sigma_{33} \rangle$ smaller than one. The ratio can be high (several hundred for the parameters used) only in the strong-coupling regime of cavity QED.

Beyond the discussed Zeno blockade, another novel effect appears in Fig. \ref{fig_regimes}(b). It concerns the population in the dark state $\ket{\Psi_1^0}$ that can be higher than the population in the ground dark state $\ket{\Psi_0^0}$. Such inversion effect cannot happen in a standard two-level system, at least not in steady state, but occurs here due to the fact that the strong atom-cavity coupling forces the atom to remain in state $\ket{2}$. This state has a projection only on the entangled dark state $\ket{\Psi_1^0}$, not $\ket{\Psi_0^0}$. Thus, as the effective Hamiltonian from Eq. (\ref{Heff_weak_coupling}) primarily promotes continuous transitions between $\left|\Psi_{0}^{0}\right\rangle$ and $\left|\Psi_{1}^{0}\right\rangle$, some degree of steady-state entanglement is expected in our system. This is indeed the case: for the parameters of Fig. \ref{fig_regimes}(b) the concurrence $C$ \cite{Wootters1998}, a measure of the degree of entanglement between bipartite $2\times 2$ systems, reaches a value of $C\simeq 0.2$, thus proving that our system truly forms an atom-photon molecule.

Additionally, for $\Omega_{23}$ comparable to $g$, the effective Hamiltonian in Eq. (\ref{Heff_RWA}) predicts non-negligible transitions to higher dark states where all decay rates $\Gamma_n^0$ are of the same order ($\Gamma_{n=1,2,3...}^0\sim \kappa$). Thus, as we see in Fig. \ref{fig_regimes}(b), appreciable population accumulates in the higher dark states.


\textit{Strong ground-state coupling regime}. The regime $\Omega_{12}\gg (g,\Omega_{23})>\kappa$ is characterized by a strong splitting of the ground states $\left|+\right\rangle$ and $\left|-\right\rangle$, as shown in Fig. \ref{fig_lvlscheme}(b). This leads to an off-resonant interaction of the atom with the laser and the cavity field. Thus, we basically observe two distinct dynamics happening simultaneously, each one describing an out-of-resonance driven two-level atom coupled to a cavity mode. Although the cavity mode and the control field oscillate at very distinct frequencies and couple different atomic transitions, in the new basis both fields couple the same transitions, from $\left|+\right\rangle$ or $\left|-\right\rangle$ to $\left|3\right\rangle$. In this case we can also derive an effective Hamiltonian, following \cite{James2007}, resulting in (see Supplemental Material\cite{supplement}  for details)
\begin{equation}
\begin{split}
H_{\text{eff}}\,{\simeq}-\frac{1}{2\Omega_{12}}\left[g^{2}a^{\dagger}a+\Omega_{23}^{2}+\left(g\Omega_{23}a+\text{H.c.}\right)\right]\sigma_{++} \\ +\frac{1}{2\Omega_{12}}\left[g^{2}a^{\dagger}a+\Omega_{23}^{2}-\left(g\Omega_{23}a+\text{H.c.}\right)\right]\sigma_{--} .
\end{split}
\label{Heffective2}
\end{equation}

This means that we have an effective nonresonant coherent drive on the cavity mode either if the atom is in the state $\left|+\right\rangle$ or in the state  $\left|-\right\rangle$, with the same probability. This can be made clearer as follows. Applying the unitary transformation $U_T(t)=\exp\left[ i\delta t a^{\dagger}a\left( \sigma_{++} - \sigma_{--}\right)\right]$,
with $\delta=g^{2}/2\Omega_{12}$, the Hamiltonian above becomes 
$H_{T} {\simeq} -\theta\left(\sigma_{++}-\sigma_{--}\right) 
- \left(\lambda ae^{i\delta t}+\text{h.c.}\right)\sigma_{++}-\left(\lambda ae^{-i\delta t}+\text{h.c.}\right)\sigma_{--},$ 
with $\theta=\Omega_{23}^{2}/2\Omega_{12}$ and $\lambda=g\Omega_{23}/2\Omega_{12}$. The first term in this Hamiltonian introduces positive (negative) time-dependent phases in the atomic states when the atom is prepared in the $\ket{+}$ ($\ket{-}$) state. The second and third terms describe nonresonant driving processes on the cavity mode, with detuning $\delta$ ($-\delta$) if the atom is prepared in the state $\ket{+}$ ($\ket{-}$). Both nonresonant processes generate a coherent field inside the cavity, but with an oscillating amplitude  and rotating in opposite directions in phase space. Thus, by properly choosing the initial atomic state and the interaction time, it is possible to generate a superposition of coherent states with opposite phases. For instance, considering the ideal situation, i.e., without cavity decay, by preparing the system in the state $\left| \Psi(0) \right\rangle =\left| 1,0 \right\rangle = \frac{1}{\sqrt{2}}\left(\left|+\right\rangle+\left|-\right\rangle \right)\ket{0}$, its state at time $t$ will be (see Supplemntal Material \cite{supplement}  for details)
$
\left| \Psi(t) \right\rangle = U(t) \left| \Psi(0) \right\rangle =\frac{1}{\sqrt{2}}\left( e^{i\phi}\ket{+}\ket{\alpha_{+}} + e^{-i\phi}\ket{-}\ket{\alpha_{-}} \right),
$
with $U(t)$ being the evolution operator, $\phi = \Omega_{23}^2 t/2\Omega_{12}$, and $\alpha_{\pm} = \pm  \Omega_{23}/g \left( e^{\mp ig^2 t/2\Omega_{12}}-1 \right)$. Then, by measuring the atom in the basis $\{\ket{1},\ket{2}\}$ one projects the cavity mode into the "Schr\"odinger-cat" states $(e^{i\phi}\ket{\alpha_{+}} \pm e^{-i\phi}\ket{\alpha_{-}})$ with $+$ ($-$) referring to detection of the atom in the state $\ket{1}$ ($\ket{2}$).


Including the decay rate of the cavity mode, both nonresonant coherent drivings present in the $H_{\text{eff}}$ will generate coherent steady states in the cavity mode, although with different amplitudes. Because of the cavity decay, the steady state of the system will be a complete mixture of the states $\left|+\right\rangle $ and $\left|-\right\rangle $. By tracing over the atomic variables, the final steady state of the cavity mode will be given by $
\rho_{ss}=\frac{1}{2}\left(\left|\beta_{+}\right\rangle \left\langle \beta_{+}\right|+\left|\beta_{-}\right\rangle \left\langle \beta_{-}\right|\right)
$, with $\beta_{\pm}=-i\frac{g\Omega_{23}}{\pm ig^{2}-2\Omega_{12}\kappa}$. For very small $\kappa$ such that $g^{2}\gg2\Omega_{12}\kappa$, we have $\beta_{+}\simeq-\beta_{-}=-\Omega_{23}/g$, which means a mixture of two coherent states completely out of phase. On the other hand, for $g^{2}\ll2\Omega_{12}\kappa$, both coherent states will have the same amplitude and phase $\beta_{+}\simeq\beta_{-}\simeq i\frac{g\Omega_{23}}{2\Omega_{12}\kappa}$. Thus, the steady state will be a perfect coherent state. This is what we see in Fig. \ref{fig_regimes}: for $\Omega_{12} > g$ the correlation function $g^{(2)}(0)$ reaches 1 (coherent state). It is important to emphasize that, for this regime, the coherent state amplitude $\beta$ scales inversely with the drive strength $\Omega_{12}$. Therefore, stronger driving will actually reduce the coherent field, indicating that this is not merely a cavity-filling effect. It is also worth noting that this coherent field is generated in the cavity mode, with a specific frequency, by driving the atomic system with laser fields with distinct frequencies. 

In order to test whether the presented scheme is experimentally feasible, we performed extensive simulations with a complete set of currently achievable parameters including the full atomic level scheme of rubidium as an example. As further discussed in the Supplemental Material\cite{supplement} , we find a behaviour that is quantitatively different but qualitatively similar to the one predicted by the idealistic model. For example, the value of almost 400 for $\langle a^{\dagger}a \rangle/\langle\sigma_{33}\rangle$ as discussed in the context of Fig. \ref{fig_regimes}d reduces to just below 100. Although experimental imperfections decrease the excited-state suppression by a factor of about four, the physical effects are robust and should be clearly visible in a realistic experiment.

To conclude, we have shown how continuous driving of a cycle between three internal atomic states can generate quantum light via a parametric wave-mixing process. The effect is experimentally robust but requires a combination of strong atom-photon coupling to a cavity and electromagnetically induced transparency by a cavity. Light emission and absorption result from transitions between dark states that are superpositions of atomic ground states entangled with cavity photon-number states. The ground-state emission suppresses decoherence due to the finite lifetime and the decay of the excited atomic state to uncoupled states. This effect is not only valuable in experiments with natural multi-level atoms but could also be beneficial for artificial atoms like quantum dots that reach strong emitter-cavity coupling but suffer from a short-lived excited state \cite{Lodahl2015}. In the near future we will investigate the effects described here experimentally, and also plan to extend the scheme by driving the ground-state transition with a laser and a vacuum cavity in mutual resonance with a second excited state. This results in a double cavity QED \cite{Hamsen2018} and double cavity EIT situation that could be employed to generate three-particle entangled states between one atom and two light fields. The atom then serves as a quantum nonlinear medium with properties that are radically different from those of a classical three-wave mixing crystal.


\begin{acknowledgments}
C.J.V.-B. thanks the support by the S\~{a}o Paulo Research Foundation (FAPESP) Grants No.~2013/04162-5 and No.~2018/22402-7, National Council for Scientific and Technological Development (CNPq) Grant No.~307077/2018-7, and Brazilian National Institute of Science and Technology for Quantum Information (INCT-IQ) Grant No.~465469/2014-0. K.N.T. was supported by the Deutsche Forschungsgemeinschaft (DFG, German Research Foundation) under Germany’s Excellence Strategy – EXC-2111 – 390814868, and B.W. was supported by Elitenetzwerk Bayern (ENB) through the doctoral program ExQM.

\end{acknowledgments}
\bibliographystyle{apsrev4-1}
%
\end{document}